# Identifying Intended Effects with Causal Models

Dario Compagno, Université de Nanterre

## Abstract

The aim of this paper is to extend the framework of causal inference, in particular as it has been developed by Judea Pearl, in order to model actions and identify their intended effects, in the direction opened by Elisabeth Anscombe. We show how intentions can be inferred from a causal model and its implied correlations observable in data. The paper defines confounding effects as the reasons why teleological inference may fail and introduces interference as a way to control for them. The "fundamental problem" of teleological inference is presented, explaining why causal analysis needs an extension in order to take intentions into account.

## Motivation: causes and their intended effects

Causal inference is about learning relationships among events (Pearl 2009). If I put a coin into a coffee machine, as a result I will obtain a cup of espresso. This is the kind of explanation offered by causal models: the coin-event works as a cause of the coffee-event. But what if I wanted to understand the reasons why somebody puts a coin into the machine, that is, for what ends some action is realized? To know the effects of some event is not enough to characterize it as action. In fact, an indefinite number of effects follow from any event, and not all of them are intended. If I put a coin in the coffee-machine, I get coffee, but I also become one coin poorer, and I waste some resources (water, electricity). It would be a mistake to see all those effects as reasons for action: I do not intentionally aim at becoming poorer, nor at wasting resources.

Explaining actions needs to answer a particular kind of counterfactual question about their reasons: if the machine did not waste resources, would the agent keep putting coins into it? (probably yes); if the machine did not produce coffee, would the agent keep putting coins into it? (definitely no). There is a need to learn how to differentiate intended and unintended effects from empirical data, starting from a causal model and hypotheses of intentional behavior. Identifying intended effects can then be done by controlling for some of the action's effects–and not for its causes, as it is done in causal inference.

## Causal and teleological inferences

We perform causal inference whenever we try to understand whether or not *A* is the cause of *B*, for example, whether the pins fall down *because* of the ball hitting them. Teleological inference is performed instead whenever we try to understand whether or not *B* is the end of *A*, for example, whether the ball is thrown *in order to* make the pins fall down. This distinction has been most clearly expressed by G.E.M. Anscombe in her book *Intention* (1957)[1].

---

[1] Anscombe has opened the way for the analytical philosophy of action and intention (Davidson 1963, von Wright 1971, Bratman 1987, Dennett 1989; see Setiya 2018, Glasscock and Tenenbaum 2023).

Both causal and teleological inferences demand to go beyond observation and data. It is not possible to observe causation: only correlations among events can be observed and registered, while causality is a category that the viewer has to add to the data in order to produce explanations of what happens. I can see that whenever the sun is high in the sky the temperature is higher, but the causal connection transcends what can be measured. The formulation of an hypothesis is needed to turn observations into evidence for or against it. This is the main idea of contemporary causal research, which developed a probabilistic and counterfactual conception of causality, discovering ways to validate causal statements with models and data (McElreath 2020, Hernan and Robins 2024).

The ends of actions cannot be observed either. One can see somebody launching a ball, and the fact that the ball hits the pins, but to conclude that the agent aimed at hitting the pins one must add *two main assumptions*: first, that there is a causal connection between the ball and the pins, and second, that hitting the pin was among the effects intended by the agent. Instead of conceiving intentions as something detached from causes, we suggest modeling action by grounding it into the causal world. This is why both causal and teleological inferences require the use of causal models. However, intentions are not causes, and cannot be modeled as such. A further layer has to be built on top of causal models to use them as tools for teleological inference.

A causal model is a set of assumptions about how the value of a variable depends on the value of other variables (its causes). A causal model can be represented as a non-cyclic graph in which nodes are variables and arcs are causal links between them. A causal model formalizes causal hypotheses and sometimes even permits to validate them with observational data (*i.e.*, without performing randomized experiments): if my model assumes that when the pins are hit by a ball they should fall, and I observe a ball hitting some pins, but they stay up, this works as evidence of the fact that my model is incorrect. Thanks to causal models, observational data become evidence for or against hypotheses, producing causal knowledge.

Causal inference is about events, while teleological inference is about actions. Both concepts of event and action can be defined with reference to causal models. An *event* is the assignment of a value to a variable $A$ which "listens" to all other variables linked to it in a causal model. The listening metaphor has been coined by Pearl (2009) as a way to express the primitive notion of causation. With reference to a causal graph, the value of $A$ depends on the values of all the variables having arcs going into $A$. If our causal graph had two variables $C$ and $A$, one for the ball and one for the pins, an arrow could connect the first to the second and so express the causal connection between the two, and more precisely that the status of the pins depends on that of the ball. So if we observe "ball thrown" ($C = c$), we expect to also observe "pins down" ($A = a$). Figure 1 shows this very simple causal graph.

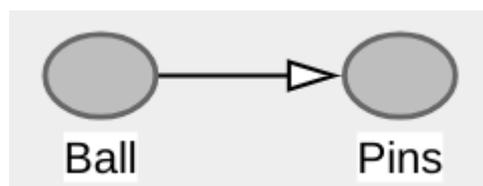

Figure 1: A very simple causal graph

---

Intentions and teleological inference are core concepts also for psychology, found in classics such as Michotte 1946 and Heider 1958 as well as in recent research like Tomasello 2022.

Events are however not enough to ground teleological explanations. Even if I knew that every time the ball is thrown the pins go down, I still do not know whether the aim of the launch was to make a strike, or some other of the effects following the launch (making noise, for example, or winning a prize). We need to define actions, and we want to do it by building a further layer on top of causal models. The reason is that representing actions as a further description of observable events anchors them to the physical world, and allows to validate teleological inference with experimental or observational data.

An *action* can be defined in this fashion as the assignment of a value *do(C = c)* to a variable *C* which "listens" to some of the variables to which it is linked to. With reference to a causal graph including *C* as a variable, the value of *do(C)* depends on the values of some of the variables having arcs coming from *do(C)*. In order to perform teleological inference, we need to tag one variable with *do()* in a causal graph: this is the variable that we consider an action and whose intended effects we want to study. Any variable *A* to which *do(C)* "listens to" is tagged in the graph as *intend(A)*. Describing the fact of throwing a ball as an action means to see it as an intervention, "pulled by" some effects identified in the model itself, such as the event of making pins fall down. Figure 2 shows a causal graph whose nodes have been tagged to express a teleological hypothesis.

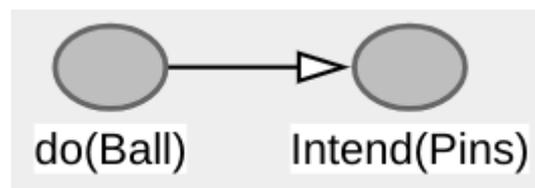

Figure 2: A very simple causal graph with nodes tagged as an action and its intended effect

The concept of intervention has been developed by Pearl as a way to validate causal hypotheses, and lays at the core of causal inference. If I want to test whether it is the ball that makes the pins fall, I should artificially isolate the ball launch from all its causes: throw it at random times and then observe whether all the times that the ball is thrown, the pins go down. In order to validate causal inferences, one needs to study the potential cause *X* independently of what made it happen, that is of its own causes. Interventions *do(X)* are a formal way to assume the free assignment of values to a variable *X*, and are needed to study its effects in an unbiased fashion.

Pearl expresses well the idea that actions are not events: observing that a ball is in motion is not the same thing as throwing the ball. Similarly, Anscombe (1957) wrote that raising my arm is not the same thing as watching my arm rise, and the difference between the two English verbs "raise" and "rise" captures the distinction between describing something as an action or instead as an event. Actions require that the agent has control over the causes of his or her behavior; an action can be seen as oriented towards an end only if we isolate it momentarily from its causes. There are an indefinite number of causes at work for me to throw a ball here and now, but what is most interesting is that if there are no pins in front of me, I do not throw.

## A counterfactual definition of intentions

Causes and ends are both concepts which need counterfactual definition, in the fashion initiated by David Lewis (1973). Limiting ourselves to facts, to what actually has or hasn't happened, doesn't give us access to their causes and reasons. Contemporary causal research has shown that causal explanations require to go beyond facts and data, and to make assumptions about what didn't happen, but could (or could not) have happened. That *C* is a cause of *A* can be formulated by saying "Had *C* = *c* not happened, then *A* = *a* would not have happened." Had the ball not been thrown, the pins would not have fallen down. This counterfactual statement expresses causality as a relationship necessarily going beyond mere observation of the actual world.

A counterfactual definition of intentions can shed new light on this concept too and ground its operationalisation. "Was *A* = *a* not an intended effect of *do(C = c)*, then *do(C = c)* would not have happened" expresses the idea that the observable event *A* is an action "listening" to at least one among its effects, namely *a*, which means that the action was done in order to produce this effect. Had the falling of the pins not been the end of the throw, the ball would not have been thrown.

To appreciate the difference between causal and teleological explanations, it is crucial to grasp that not all the effects of an action are intended. Causal inference could ideally manage to identify all the effects of an event *C* = *c*, and still give absolutely no information about which of these effects was the reason for an agent to make this event happen. It is also important to understand that the "listening" metaphor coined by Pearl has to be interpreted differently with reference to causal or instead teleological explanation.

Let us imagine a stove and a pot of water on it. The aim of causal inference is to identify a causal model (*i.e.*, to put it to the test with respect to empirical data). For example, the model expressing the idea that "Water is boiling because the stove is hot" formalizes that the variable for water listens to the variable for stove, and so that if *stove* = *hot*, then *water* = *boiling*. This is as far as causal inference goes. The aim of teleological inference is instead to identify a means-ends relationship within a causal model, that is, an association between an action and some intended effects. The aim is therefore to formalize and put to the test an assertion like "The stove is hot in order to make water boil."

It would be a mistake to think that actions operate under some sort of backwards causation (see Faye 2001). The stove does not "listen" to the water (hot water does not turn the stove on by itself), causation goes in one direction and cannot be reversed. However, the observation *stove* = *hot* and the action *do(stove = hot)*, turning the stove on and so assigning the value *hot* to the variable *stove*, are not the same thing and have to be treated differently, as Pearl has shown. It is exactly the difference between a variable *X* and an intervention *do(X)* over it, which we use here to model actions on top of causal assumptions.

In fact, an intervention *do(X)* cannot be represented within a causal model, as it does not stand for a variable *X* but for an operation on the causal model including the variable *X*, arbitrarily assigning a value to it. If I turn the stove on arbitrarily, at random times, I am isolating the stove from its causes, *i.e.* from the other events determining its on or off status within a causal system (such as the times of lunch and dinner, the day of the week, etc.). This is why it is correct to write that *do(stove)* "*listens*" *to water*. The action affecting the stove is done in order to have a causal impact on the water, respecting the direction going from cause (the stove) to effect (water).

We are suggesting therefore to give two different interpretations to the "listening" metaphor, the first causal (proposed by Pearl), and the second teleological (inspired by Anscombe). Events "listen" to their causes, while actions "listen" to some of their effects. This "listening" relationship is anyway the aim of modeling and explanation.

We clarified that *do(stove)* "*listens*" *to water* is not the same thing as *stove* "*listens*" *to water*. It is important to notice that interpreting actions depends on the correct recognition of a grounding causal relationship, in this case that *water listens to stove.* It is thanks to the causal assumptions grounding action that intention can be identified from causal models and data. Starting with a causal model which describes the relationships of cause and effect among variables, then observable correlation can become evidence for or against teleological hypotheses. The agent turns on the stove *only if* she believes that her action will have the effect of making the water boil. And her intention can be understood by us *only if* we refer to the same causal model, in which the stove is the cause and water the effect. In fact, if the agent didn't want to boil water, she would not have turned the stove on, independently from the other effects of her behavior. This last sentence is what needs to be operationalized to validate teleological inferences.

## Confounding effects and the "fundamental problem" of teleological inference

Confounding causes are what has to be excluded by causal inference in order to identify causal models. In fact, some observable correlation between two variables *A* and *B* may not be due to a causal relationship between the two, but to a third variable *C* which is a common cause of both (Reichenbach 1956). For example, I may observe that education and salary are positively correlated, and guess that the first is a cause of the second, but actually there is a third variable (let's say family status) which causes both education and salary, leaving little variation still to be explained.

If I artificially cancel out the impact of family status on education, I am assigning values to education independently of its causes, as if I was performing the action *do(education)* and observing whether or not my value assignments are correlated with the observable values of the variable salary. In fact, in order to explain the correlation of education and salary causally, I need to make sure it is not due to any common cause and so, specifically to this model, that the values of education do not depend on family status. Were I to perform an imaginary randomized experiment, I could take subjects at random from different family backgrounds and force them to diverse educational paths, observing how much they end up earning years later. This kind of action is what is called *intervention* by Pearl, and in certain conditions it can be simulated with purely observational data, without the need to perform randomized experiments. Figure 3 shows this causal graph with one variable tagged so to express that its value is assigned by an intervention.

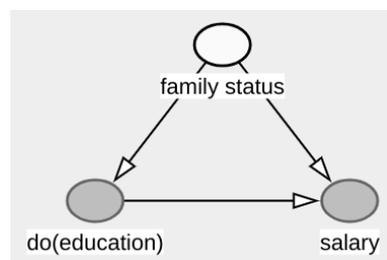

Figure 4: A causal graph with one intervention tagged with *do()*

Several strategies have been developed to control for confounding causes in observational data. For example, in order to test whether education has an effect on salary, I may isolate the impact of family by stratifying families according to their status and check in each stratum whether I keep observing a correlation between education and salary. If the correlation is still there, I can conclude that my causal hypothesis linking education to salary resists a tentative falsification made by taking into account family status. The fact that my hypothesis has endured falsification makes it more probably true (McElreath 2020); as a matter of fact, causal models are a way to make potential confounding variables explicit, in order to find ways to control for them, resulting in causal knowledge based on explicit assumptions.

Let us imagine a slightly more complex causal model, and let's assume that this model represents well (to the best of our knowledge) the cause-effect relationships among its variables. The variables in the model are *sport practice*, which causes one to *lose weight*, which in turn causes one to *be fit*, which finally causes one to *live longer*; there is also a fifth variable *win medals* which is directly caused by *sport practice*. On the basis of this causal model, we can formulate several teleological hypotheses, among which the one asserting that *sport practice* is done in order to *be fit*. Let us show our teleological hypothesis on a causal graph by tagging sport practice with the *do()* tag, and *be fit* with the *intend()* tag. This latter operator is used to highlight those effects that we believe are intended by the agent when performing the action tagged with *do()*. The tagged causal graph is shown in Figure 4.

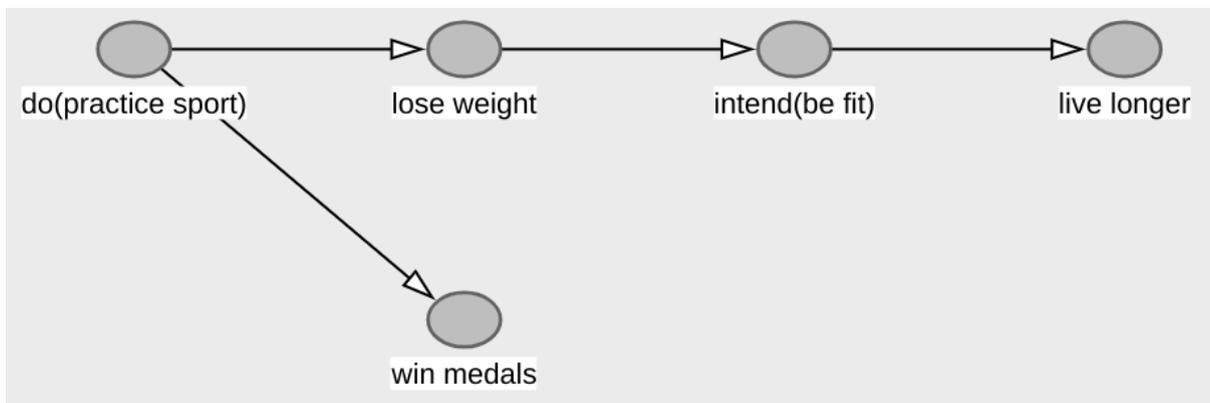

Figure 4: A more complex causal graph tagged with *do()* and *intend()*

Inferring intentions is somehow harder than inferring causes because of something that may be called the "fundamental problem" of teleological inference: *among all the effects caused by the event* A, *only some are intended by the action* do(A). So even if we knew all the correct causal relationships involving the variable *A*, it would still not be possible to guess which effects of *do(A)* are intended, given that all the effects of *A* co-occur. Practicing sport makes you lose weight, and at the same time makes you fit, and at the same time makes you live longer, and at the same time makes you win medals. Given that all these effects follow from the same cause, how is it possible to discern the ones which were intended by the agent?

Whenever one makes a teleological inference about somebody else's intention, there is the risk of missing the intended effects, or of mistaking other effects for intended. The effects which co-occur with the intended ones, but are not intended, could be called *confounding effects*, as they are the reason why teleological inference may fail. For example, I may make the hypothesis that agents practice sport in order to be fit, but instead they actually do it in order to win medals. The aim of model based teleological inference is to exclude confounding effects and identify a means-ends relationship, for example that the intended effect of practicing sport is to be fit, despite the fact that sport makes you also win medals.

If we go back to Figure 4 and our model about the effects of sport practice, we can observe that there are three kinds of potential confounding effects. Let us remember that our hypothesis is that sport is done in order to be fit, so the variable *be fit* is the intended effect we are aiming to identify. First, there are *further effects* that may be the actually intended ones: people may practice sport in order to live longer, and not really in order to be fit. Our teleological hypothesis would be "too short," so to speak, and it should have aimed further in the causal chain of effects produced by action, as the true intended effect is a descendent in the causal graph of the one we initially guessed. Second, we should check for *mediating effects*: one may practice sport in order to lose weight, and not really to become fit. Our hypothesis about the agent's intention would therefore go "too far," and miss the real end, that lays between the action and our guessed intended effect. Third, we need to exclude *parallel effects*: sport could be done in order to win medals, independently of the fact that athletes also lose weight, are fit and live longer. The effect *be fit* that we thought was the intended one, would lay in a different branch of the causal graph with respect to the real one.

## Interfering with confounding effects

How can we control for confounding effects, and become more confident of our teleological hypothesis? What we need is a special kind of intervention, which could be called *interference*. In causal inference, intervention is about controlling for potential confounding causes; interference is instead a way to control for potential confounding effects. How can I be sure that sport is not practiced in order to win medals? For example, I can forbid athletes to enroll in competitions, and then observe whether they keep practicing sport anyway. If they do, with the same frequency as before my interference, this is evidence that winning medals wasn't one of the reasons for sport practice; otherwise, this shows that the action *do(practice sport)* was actually listening to its effect *winning medals*, and so that winning medals should be accounted among the reasons for action.

Interference is a sort of intervention based on the assumption that one variable in the causal model is an action "listening to" some of its effects; the effects of such action are interfered with, and then the values of the action are observed. If I prevent athletes from winning medals, I expect to observe a significant reduction in their sport practice *only if* I assume that practicing sport (cause) is an action oriented towards the end of winning medals (effect). This clearly goes beyond causal inference. In fact, an ordinary intervention would prevent people from practicing sport (cause) and observe whether they win less medals (effect). This would provide evidence for the causal relationship between sport and medals, but would say nothing about the ends of sport practice.

Interference can be realized via a randomized experiment. I could separate athletes in two groups and forbid the first to enroll in competitions. Any significant change in sport practice between the two groups can then be interpreted as evidence for the teleological hypothesis that sport is done in order to win medals. Is this a causal experiment? It cannot be: as a matter of fact, practicing sport makes you win medals, and not the other way around. An honest causal experiment would discover very quickly that people who do not practice sport never win medals. Instead, we aim at controlling experimentally for one of the effects of practicing sport, which is definitely not how causal research should be conducted.

One may think that somehow the relationship going from winning medals to practicing sport, if real, *must* be causal, but this is bad reasoning. Such fallacy becomes immediately evident if we go back to our simpler causal model in which hot stoves cause the pots of water on them to boil: there is no way that boiling water turns the stoves on by itself; and still, if I prevent people from obtaining cold water to cook with (or if I isolate thermally the pot from the stove), they will not turn the stoves on for no reason, and I will observe that the stoves remain off. So I can experimentally produce a significant impact in a cause (stoves) by interfering with some of its effects (water), as long as the interfered effect was really among the ones intended by the agent when intervening on the cause. This is how teleological inference can be identified starting from a causal model and data.

Can we control for the effects of an action without realizing randomized experiments? Observational data can become evidence for teleological inference under certain conditions. In our example, we need data about people who enroll in competitions and people who do not. We also need to understand which variables have to be balanced (see Hernan and Robins 2024) so as to simulate most exactly as possible the results of a randomized experiment.

In fact, what we need evidence for is the relationships between *do(practice sport)* and *win medals*. Therefore, we have to assume that sport practice is an action, and we should make sure that the data collected about *sport practice* approximate those that could be collected about *do(sport practice)*. Practically, this means to control for all the causes of *sport practice* in our causal model, which are also causes of winning medals. For example, it may very well be that *age* is a confounding cause of practicing sport and of winning medals; any inference would therefore be biased if it didn't take this information into account and control for it. However, as long as the values of practicing sport are independent from its causes (that is, approximately assigned at random), interfering with its effects has teleological value.

We stated above that there are three kinds of potential confounding effects: further effects, mediating effects and parallel effects. The model shown in Figure 5 exemplifies the three kinds and includes the variables about which we need data as evidence for or against the teleological hypothesis that sport is done in order to be fit and so that *do(practice sport)* "*listens to*" *be fit*.

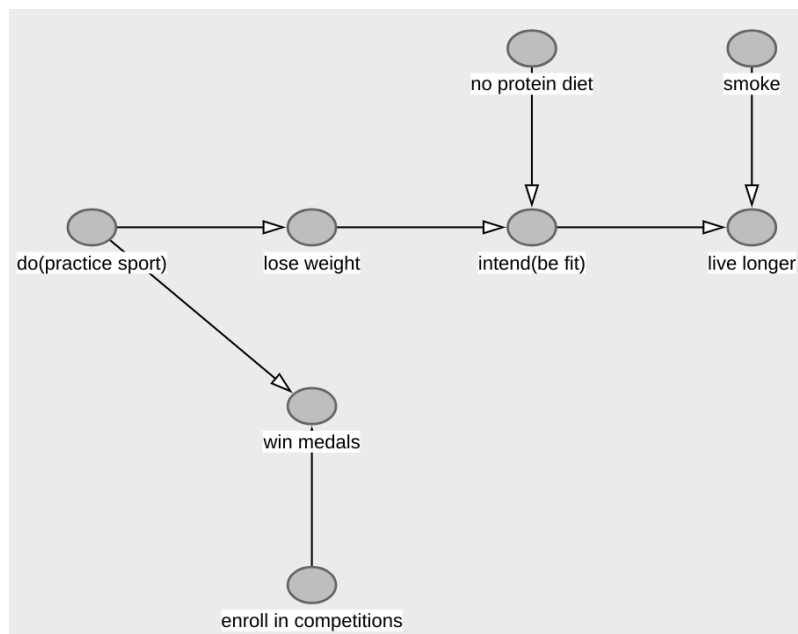

Figure 5: Variables needed to control for the three potential confounding effects in the model

Parallel effects and further effects are controlled for in an analogous fashion: we need to interfere with them and observe if the agents keep performing the action with the same frequency. In our example, we may forbid the persons in our sample to enroll in competitions, so as to interfere with *win medals*, and we may force them to smoke, so as to interfere with *live longer*. If our hypothesis *intend(be fit)* is correct, we expect no significant change after our interference. Tables 1 and 2 show the data that we need to observe as evidence for our hypothesis; the variables are simplified as binary variables which can only take the values 0 or 1.

| Practice sport | Lose weight | Be fit | Win medals | Enroll in competitions |
|---|---|---|---|---|
| 1 | 1 | 1 | 0 | 0 |

Table 1: What we need observing to exclude *win medals* as potential confounding effect

| Practice sport | Lose weight | Be fit | Live longer | Smoke |
|---|---|---|---|---|
| 1 | 1 | 1 | 0 | 1 |

Table 2: What we need observing to exclude *live longer* as potential confounding effect

If we observe that *enroll in competition = 0* co-occurs with *practice sport = 1*, our teleological hypothesis that the intended effect of practicing sport is to be fit grows stronger, because we can exclude that people do sport for winning medals; similarly, if we observe together *smoke = 1* and *practice sport = 1*, we can exclude that sport is done for living longer.

Controlling for mediating effects is a little more subtle, but not more complicated. In fact, we cannot control for a mediator, as this would result in controlling also for its effects, among which lays the one we believe is intended (see Pearl, Glymour and Jewell 2016). If I prevent people from losing weight, they cannot become fit either, given that losing weight is the way by which people become fit according to our simple causal model. However, we can control for the intended effect itself: if after doing it we observe any significant change in the action, we have proof that our hypothesis was correct; otherwise, we know that the supposedly intended effect wasn't the actual end of action, and we should look at effects which mediate between it and the action. We could prevent people from getting fit by imposing a strict no protein diet; this means that people who practice sport still lose weight, but they do not become fit anymore. If our hypothesis *intend(be fit)* is correct, we expect people to stop practicing sport under these conditions. If instead they keep practicing it, despite our interference, this means that *be fit* and all of its descendents in the causal graph weren't the actual reason for action. Table 3 shows the data that we need *not* to observe if our hypothesis is correct.

| Practice sport | Lose weight | Be fit | No protein diet |
|---|---|---|---|
| 1 | 1 | 0 | 0 |

Table 3: What we need *not* observing to exclude *lose weight* as potential confounding effect

# Conclusion

Intentions play an important role within the human, social and psychological sciences–possibly a role which is as important as that of causes. Until now however this concept lacked operationalization. We have shown that intentions can be inferred from a causal model and data, if we define them counterfactually within a probabilistic framework. Intentions, like causes, cannot be seen or measured, but a counterfactual approach and causal models are able to turn data into causal or teleological evidence.

Humans perform teleological inferences all the time: we simply know if the person next to us has a knife in his hand to spread the jam or instead to draw blood. We know this, not because we magically get into other people's minds, but because we observe the effects of their actions and understand how they control for them. Still, our intuitive teleological inferences are like our intuitive causal inferences: they may very well be biased. Causal research has opened a way to operationalize a probabilistic and counterfactual concept of causality, which permits today to express and validate precise causal statements from data. The present work aims at opening an analogous possibility for teleological inference, based on the framework and results of causal research, but going beyond them as intentions are not causes and cannot be modeled as such.

Causal research already makes use of interventions, which can be easily extended to inquiry about their intended and unintended effects. What we call the "fundamental problem" of teleological inference shows why causal and teleological inference are necessarily distinct, and the present paper aims to ground strategies to overcome this problem. Particularly, we have seen how to control for the three kinds of potential confounding effects which may make teleological inference fail.